# 3D-QCNet - A Pipeline for Automated Artifact Detection in Diffusion MRI images

Adnan Ahmad[1], Drew Parker[1], Zahra Riahi Samani[1], Ragini Verma[1]

[1]Diffusion and Connectomics in Precision Healthcare Research Lab (DiCIPHR), Department of Radiology, University of Pennsylvania, Philadelphia, PA

Correspondence: Ragini Verma
ragini@pennmedicine.upenn.edu

## Abstract

Artifacts are a common occurrence in Diffusion MRI (dMRI) scans. Identifying and removing them is essential to ensure the accuracy and viability of any post processing carried out on these scans. This makes QC (quality control) a crucial first step prior to any analysis of dMRI data. Several QC methods for artifact detection exist, however they suffer from problems like requiring manual intervention and the inability to generalize across different artifacts and datasets. In this paper, we propose an automated deep learning (DL) pipeline that utilizes a 3D-Densenet architecture to train a model on diffusion volumes for automatic artifact detection. Our method is applied on a vast dataset consisting of 9000 volumes sourced from 7 large clinical datasets. These datasets comprise scans from multiple scanners with different gradient directions, high and low b values, single shell and multi shell acquisitions. Additionally, they represent diverse subject demographics like the presence or absence of pathologies. Our QC method is found to accurately generalize across this heterogenous data by correctly detecting 92% artifacts on average across our test set. This consistent performance over diverse datasets underlines the generalizability of our method, which currently is a significant barrier hindering the widespread adoption of automated QC techniques. For these reasons, we believe that 3D-QCNet can be integrated in diffusion pipelines to effectively automate the arduous and time-intensive process of artifact detection.

Keywords: Quality Control, Artifacts, MRI, Diffusion MRI, Deep Learning

# 1. Introduction

Diffusion MRI (dMRI) has become a widely adopted imaging technique over the past few years [1], as it is able to provide a unique insight to white matter architecture by exploiting the differential water diffusion across tissue type. dMRI is used widely to study conditions where WM abnormalities are expected such as brain tumors, brain trauma, neurodegenerative and neuropsychiatric disorders [3]. However, dMRI acquisition suffers from artifacts that can make their use in research studies challenging in the absence of intensive quality control (QC) [4-6]. Thus, our goal is to develop a deep learning based pipeline that will automatically detect and isolate artifact ridden volumes from dMRI scans dMRI artifacts usually include motion induced signal drop-out, ghosting, herringbone, Gibbs ringing, chemical shift, susceptibility, interslice instability and multi-band interleaving [7-10] [11-13]. Patient motion during the scan can result in signal dropout from individual slices in a volume, in ghosting artifact, or in artifacts arising from stitching together misaligned data in the cases of acquisitions using slice interleaving or multi-band acceleration. Field inhomogeneity from interfaces between water and fat or between tissue and air (as in the sinuses), or from metal implants, can cause distortions such as chemical shift and susceptibility artifacts.

The detrimental effects of using scans marred by artifacts on further post-processing has been noted extensively and have been shown to negatively affect the viability of results [14-16], and also affect their interpretation. This makes it imperative to detect these artifacts so that the artifactual data can either be excluded from further analysis or corrected as part of a preprocessing pipeline. Hence QC for the detection and removal of these artifacts is an essential part of a dMRI processing pipeline [3].

Currently, most QC happens through manual intervention, where an expert manually goes through all scans, flagging those that they deem abnormal. This is an arduous and time intensive process. It can be infeasible when large datasets are involved and quickly becomes a bottleneck for any dMRI study. Further, the unavoidable subjectiveness of having different experts annotate different studies will lead to uncertain outcomes since two individuals may have a varied opinion of what may constitute an artifact. This makes a normalized and automated QC method as we propose, which can replicate or exceed human performance, imperative for any dMRI processing pipeline. While some form of QC tools already exist like FSL EDDY [17, 18], DTI Studio [19], DTI Prep [20] and TORTOISE [21], they are usually limited to detecting and correcting the specific artifact they have been designed for, mostly motion and eddy current induced distortions [22-26]. Moreover, they are not entirely accurate, as often the results of their correction need subsequent inspection to detect the presence of any remaining abnormalities. These methods are also not exhaustive in covering many other prevalent artifacts like ghosting, herringbone and chemical shifts. Furthermore, Deep learning (DL) methods like QC Automator [27] have shown excellent performance in detecting a wide range of artifacts. The QC automator trained and tested a CNN classifier using transfer learning and finetuning to effectively classify artifacts from 2D slices in both the axial and sagittal views. This method further demonstrated the superior performance that deep learning based methods have over more conventional learning algorithms when it comes to artifact detection.

However, despite recent success there still remain a few factors that prevent automated methods from being effectively integrated into diffusion pre-processing pipelines. First and most significantly, no previous method has been validated on data containing pathology which is a very common use case in a practical scenario. Second, 2D deep learning models like QC-Automator, require ground-truth annotations for every slice from every volume which can be strenuous for human annotators. This can be particularly concerning specifically during model fine-tuning where an annotated subset is required every time an unseen dataset is to be processed. Lastly, previous studies have not addressed how to account for the subjective differences; multiple users of an automated qc pipeline may have when defining artifacts.

Our proposed method, 3D-QCNet attempts to solve these issues by implementing a 3D architecture which identifies artifacts accurately from both patient and healthy diffusion scans while simultaneously streamlining the task of annotation and optimizing the process of fine-tuning. We show that our method achieves excellent performance on a wide range of diverse datasets sourced from different scanners, patients, and pathologies. Further, unlike a 2D model our annotators only had to assign a single label to an entire volume as against to a label for every slice. This led to monumental savings in terms of both time and effort. As a result, we were able to label and train on 3 times more data than used by QC Automator in considerably less time. It also made our pipeline significantly more user friendly and practical when base models were needed to be finetuned on previously unseen datasets. Additionally, to account for the variance in sensitivity different users may have when deciding which artifactual volumes to remove, we added a flexible probability threshold which can be selected based on how specific the user desires to be in his study. Since our proposed method was planned by keeping in mind practical use cases where scans are heterogenous by nature, our intention was to make our model agnostic to a range of factors including scanner differences, pathology presence and type of artifacts. Therefore, our objective was to develop a solution to detect artifacts that will work out of the box in most practical situations with minimum overhead.

Overall, in this paper we present a user-centric automated 3D QC pipeline for DTI scans which is tested to be both accurate and flexible on data observed in most practical scenarios while being easier to fine-tune on unseen distributions. This makes it convenient to integrate in existing preprocessing dMRI pipelines, hence automating an essential but previously time consuming and laborious process.

## 2. Methods and Data

### 2.1 Overview

Our method utilized a 3D-DenseNet [28] architecture which classified dMRI volumes into an Artifact class or a Normal class. We began by first extracting 3D volumes from dMRI scans, preprocessing them to a common size and then training a deep learning model optimized to solve this classification task. We also showed how provisions like flexible probability thresholds and fine-tuning can be easily used to tweak performance.

In the following sections, we first describe the various datasets we used to train and validate 3D-QCNet. We then discuss the specifics of our method – its architecture, training process, performance optimizing techniques and finally our evaluation criteria.

## 2.2 Data

All work in this paper was carried out with the approval of the IRB of the University of Pennsylvania. In this study, we used 7 datasets of which 3 were utilized for training and validation (Dataset 1, 2 and 3) while the other 4 were used exclusively for testing (Dataset 4, 5, 6, 7) model performance on unseen data distributions. These datasets were heterogenous in nature having been sourced from scans acquired with diverse scanning parameters, having different pathologies and subject demographics. The specific details for each dataset can be found in Table 1a and Table 1b. The volumes were selected randomly for annotation from each of their respective larger datasets. This resulted in a total ground truth labelled dataset of 9258 volumes from 678 subjects. The annotations were done by an expert (DP) with 8 years of experience identifying artifacts in diffusion scans. The artifacts that were targeted by the expert included motion-induced signal dropout, interslice instability, ghosting, chemical shift and susceptibility.

*Table 1a. Dataset details - acquisition parameters*

|  | Dataset # | Original - volumes/ subjects | Annotated -volumes/ subjects | Scanner - Siemens 3T | TR \| TE (ms) | Resolution (mm) | b-value | # Dirs | Multib and Factor |
|---|---|---|---|---|---|---|---|---|---|
| Training - Validation | 1 | 10997/78 | 1263/77 | TimTrio | 6500\|84 | 2.2 x 2.2 x 2.2 | 1000 | 30 | NA |
|  | 2 | 99995/1419 | 2703/94 | Verio | 8100\|82 | 1.875 x 1.875 x 2 | 1000 | 64 | NA |
|  | 3 | 5117/165 | 2945/165 | Verio | 11000\|76 | 2 x 2 x 2 | 1000 | 30 | NA |
| Testing | 4 | 6440/85 | 1098/44 | PrismaFit | 2900\|94 | 2.4 x 2.4 x 2.4 | 1000 | 64 | 3 |
|  | 5 | 55937/633 | 400/242 | TimTrio | 8000\|82 | 2.2 x 2.2 x 2.2 | 1000 | 30 | NA |
|  | 6 | 68460/585 | 600/30 | Skyra | 9000\|92 | 2.73 x 2.73 x 2.7 | 1300 | 64 | NA |
|  | 7 | 3063/26 | 249/26 | PrismaFit | 4300\|75 | 2 x 2 x 2 | 300, 800, 2000 | 109 | 2 |

*Table 1b. Dataset details – pathology information*

|  | Dataset # | Type | Age range (y) |
|---|---|---|---|
| Training and Validation | 1 | TBI dataset with Lesions and WM hyperintensities | 18-66 |
|  | 2 | Developmental dataset | 8-22 |
|  | 3 | Autism dataset | 6-25 |
| Testing | 4 | TBI dataset with Lesions and WM hyperintensities | 18-71 |
|  | 5 | Hypertension, cardiovascular disease and WM hyperintensities | 55-94 |
|  | 6 | TBI dataset with Lesions and WM hyperintensities | 18-71 |
|  | 7 | Healthy Controls | 24-26 |

## 2.3 Data Distribution

We distributed our data into 3 sets – training, validation and testing. Datasets 1,2 and 3 were exclusively used for creating the training and validation sets. The split was done in a stratified manner such that 25% of subjects from both the Artifact and Normal class were put in the validation set and the remaining were used as part of the training set. This resulted in 5619 training volumes and 1292 validation volumes from a total of 336 subjects. It is worth noting, that we split our datasets based on subjects rather than slices, to prevent data leakage that may occur when different volumes from the same subject are present in both training and validation sets. Our testing set, which constituted 2347 volumes from 342 subjects, was entirely sourced from 4 unseen datasets (Datasets 4, 5, 6 and 7) that were not used in any way during training and validation.

## 2.4 Model and Architecture

Our method is inspired by the success that Convolutional Neural Architectures [29] have had in the domain of Computer Vision [30]. CNNs are adept at identifying spatial relationships in images by determining differentiating features of varying granularity along their depth. These automatically extracted features are then passed through a conventional dense Neural Network (NN), which is trained end to end to optimize the loss function such that the model learns to classify among the different classes.

We build upon the work done in QC Automator by adapting a deep learning model to solve a 3D classification problem. In order to accomplish this task, we use a special kind of CNN architecture called DenseNet which has been shown to have superior performance compared to other architectures in multiple classification tasks [28].

The DenseNet architecture consists of Dense blocks interconnected with each other using Transition Layers. Within each Dense block, the input is passed through BatchNorm [31] and Activation layers before being sent to multiple 3D Convolutional layers. Significantly, the output from the CNN layers is concatenated with their input creating skip connections. This feature differentiates DenseNet from other architectures like ResNet [32] where skip connections exist but concatenation is replaced by the addition operation. The outputs from the Dense block are then sent to the Transition blocks which are responsible for downsampling the feature maps, an essential operation for any CNN model. Since the concatenation operation inside Dense blocks enforces shape limitations, this operation has to be carried out outside of these blocks in Transition Cells before the output is sent to the next Dense block. These two blocks form a chain as they continue to process the feature maps in varying order of granularity along the depth of the model.

The resultant output is then sent to the Global Average Pooling layer which further reduces dimensions by averaging across feature maps. The final output can be considered as the features extracted by the CNN and are subsequently fed through a set of Dense NN layers to obtain the softmax output.

We configured our DenseNet model to have 3 Dense blocks each having 2 3D-Convolutional layers with a filter size of 3x3x3 and skip connections between their inputs and outputs. The Convolutional layers were additionally preceded by BatchNorm and Relu Activation layers. Each transition block further consisted of 4 layers comprising BatchNorm, Activation, Conv and Average Pooling. The output emanating from the last GAP layer was fed to a single Dense layer of size 2 having the softmax activation. This output was then optimized against the target distribution through the loss function. Fig. 1 describes the architecture in further detail.

We preferred using DenseNet over other architectures for a few reasons [28]. First, the skip connections allow for easier transmission of gradients during back propogation, diminishing the vanishing gradient problem. This allows for a deeper architecture enabling the model to learn features at different granular levels. Also, the number of learnable parameters in a Densenet architecture are considerably less which leads to a reduced tendency for the model to overfit on the training data.

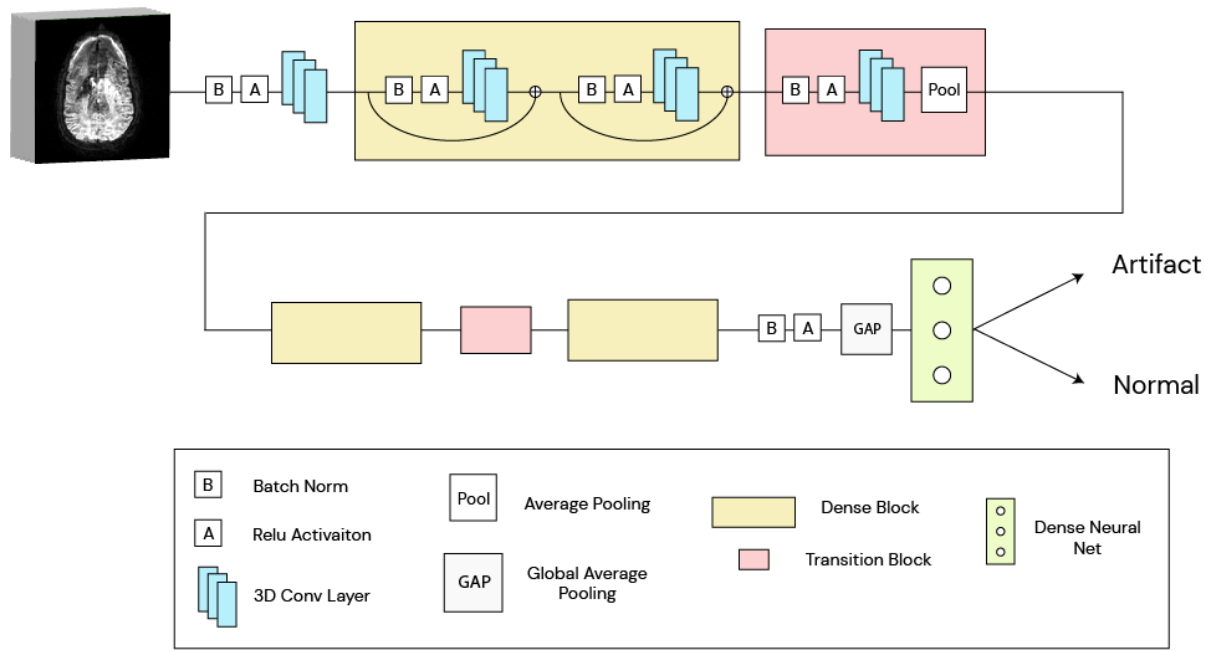

*Fig. 1 3D-QCnet DenseNet Model Architecture showing how a 3D volume is processed by a series of Dense and Transition blocks before being sent to the GAP and Dense Layers for final classification into the two classes – Artifact and Normal*

Our training setup for the 3D-QCNet model involved preprocessing in which we resized all our 3D volumes extracted from diffusion scans into a common dimension size of 96x96x70. To achieve this, we clipped volumes which had more than 70 slices while adding empty slices as padding for those which were below this threshold. Further, while training we used a batch size of 5 due to the increased memory requirements of 3D models and set the epochs to 20. The model was trained to optimize cross-entropy loss using the Adam optimizer with a learning rate of 10e-4. On a Nvidia 1080 GPU the model took approximately 3 hours to train. The code for our method was written in Python 3.6 and utilized the Tensorflow and Keras libraries for model implementation.

2.4.1 Fine-Tuning

dMRI data is known to be heterogenous in nature due to various factors like scanner differences, diverse pathologies, among others. While we trained our model on a relatively large dataset, it is impossible to exhaustively account for all these intricate differences that may be found while processing a new study. To best enable our NN model to generalize over a previously unseen data that may have different attributes to the one on which the model was trained, we allow the fine-tuning of our base model as a means to obtain the best possible results on this new target distribution. We allow users to take a small subset of the target dataset, use that to retrain all the layers of the base model for a few epochs and utilize the resultant model for further analysis. As it requires additional annotation effort, we suggest that fine-tuning be reserved for those cases where the performance of the 3D-QCNet model on an unseen dataset is well below the acceptability criterion.

### 2.4.2 Flexible Decision Threshold

Conventionally, DL methods apply 0.5 as a probability threshold on the final softmax output to categorize a datapoint among the two classes. While this is a reasonable assumption, during our analysis we found different expert annotators to have different criteria when marking a scan as artifact - some were more sensitive in discarding a scan than others. Therefore, to integrate a similar subjectiveness in the decision-making process of our model, we allowed our users to select the probability threshold based on their own sensitivity/recall and precision preferences. For instance, increasing the threshold leads to a higher bar in what it would take for a scan to be marked as an artifact, therefore only those scans which the model is extremely confident are marked as such. This leads to an increase in precision but may also simultaneously result in a decrease in recall. Changing the threshold in one direction, therefore, impacts the balance between these two metrics and the user can choose a value where they find the best tradeoff for their use case.

## 2.5 Evaluation Metrics

We evaluated the performance of our method by defining 'Artifacts' as the positive class and observing the Precision, Recall and Accuracy metrics which are defined as follows:

$$Precision = TP / (TP + FP)$$

$$Recall = TP / (TP + FN)$$

$$Accuracy = TP + TN / (TP + TN + FP + FN)$$

Here, TP, FP, FN, TN represent True Positive, False Positive, False Negative and True Negative respectively.

## 2.6 Inference Pipeline

Overall, our pipeline works in the following steps when used to curate a new dataset. First, it accepts dMRI scans, extracts 3D volumes and preprocesses them into a constant size before sending the data to the 3D-QCNet model. The model runs in inference mode and provides the user with predicted artifact/normal label for each volume. Should a test set be provided, performance metrics are calculated and reported. If through inspection of model output or test metrics, the user finds the performance lacking, they may annotate a small subset of data and fine-tune the model on it. Flexible thresholds can further be used to tweak the final precision-recall balance. At the end, the user is presented with a report stating the volumes which were found to have an artifact. The user may then choose to remove these volumes before proceeding to further downstream analysis.

## 3. Results

### 3.1 Performance of 3D-QCNet on Validation and Test Data

Table 2 below represents performance of 3D-QCNet on the validation and test sets. The model obtains excellent metric scores with an average accuracy of 87% and 95% on the validation and test sets, respectively. Significantly, a high average recall of 92% is also obtained on the test set and no huge discrepancies in performance are observed across the diverse group of 7 datasets.

*Table 2. Results – 3D-QCNet Model*

|  | Dataset | Threshold | Accuracy | Precision | Recall |
|---|---|---|---|---|---|
| Validation Set | Dataset 1 | 0.15 | 96 | 88 | 84 |
| | Dataset 2 | 0.15 | 77 | 85 | 70 |
| | Dataset 3 | 0.15 | 88 | 82 | 90 |
| | *Average* | | 87 | 85 | 81 |
| Test Set | Dataset 4 | 0.5 | 97 | 84 | 81 |
| | Dataset 5 | 0.5 | 92 | 86 | 98 |
| | Dataset 6 | 0.5 | 96 | 81 | 100 |
| | Dataset 7 | 0.5 | 92 | 95 | 89 |
| | *Average* | | 94 | 87 | 92 |

### 3.2 Fine-Tuning and Threshold improvements

Fine-tuning 3D-QCNet was not essential to improve performance on the test set since results comfortably exceeded acceptability criterions. However, it might be a useful technique to optimize on certain datasets in the future that present a wildly different distribution that the DL model does not properly generalize to. To simulate performance improvement due to fine-tuning, we trained a model on one dataset, Dataset 1, and tested the performance of this model on 2 different datasets, Dataset 3 and 4. We did this by first testing directly and then by selecting a 10% subset from the target datasets and fine-tuning the base model on it. We also use these experiments to show how different threshold values can affect the precision-recall distribution.

Table 3. Effects of Fine-tuning and Flexible Thresholds

| Dataset 4 | | Threshold | Accuracy | Precision | Recall |
|---|---|---|---|---|---|
| | Base Model | 0.5 | 77 | 100 | 30 |
| | | 0.07 | 87 | 85 | 72 |
| | Fine-Tune | 0.7 | 85 | 85 | 80 |

| Dataset 5 | | Threshold | Accuracy | Precision | Recall |
|---|---|---|---|---|---|
| | Base Model | 0.5 | 83 | 100 | 10 |
| | | 0.1 | 92 | 82 | 73 |
| | Fine-Tune | 0.5 | 93 | 80 | 82 |

## 3.3 Selecting the Ideal Fine-tuning set

Selecting a subset for fine-tuning can often involve contentious and subjective decisions regarding its size and distribution. Therefore, building on the experimental setup above, we further analyzed the ideal size of the fine-tuning subset and how selecting different sets can change performance metrics in Table 4.

Table 4. Selecting a Fine-Tuning set

| Dataset | Samples | Subset Type | Class Distribution | Thresholds | Accuracy | Precision | Recall |
|---|---|---|---|---|---|---|---|
| | | | | | | | |
| Dataset 4 | 222 Volumes/ 12 Subjects | 10% of Volumes | 32% artifacts | 0.7 | 84 | 70 | 85 |
| | 111 Volumes/ 71 subjects | 5% of Volumes | 32% artifacts | 0.3 | 88 | 82 | 81 |
| | | | | | | | |

|  | 236 Volumes/ 34 subjects | 30% of Volumes | 6% artifacts | 0.5 | 93 | 80 | 82 |
| --- | --- | --- | --- | --- | --- | --- | --- |
| Dataset 5 | 118 Volumes/ 34 subjects | 15% of Volumes | 6% artifacts | 0.5 | 93 | 83 | 75 |

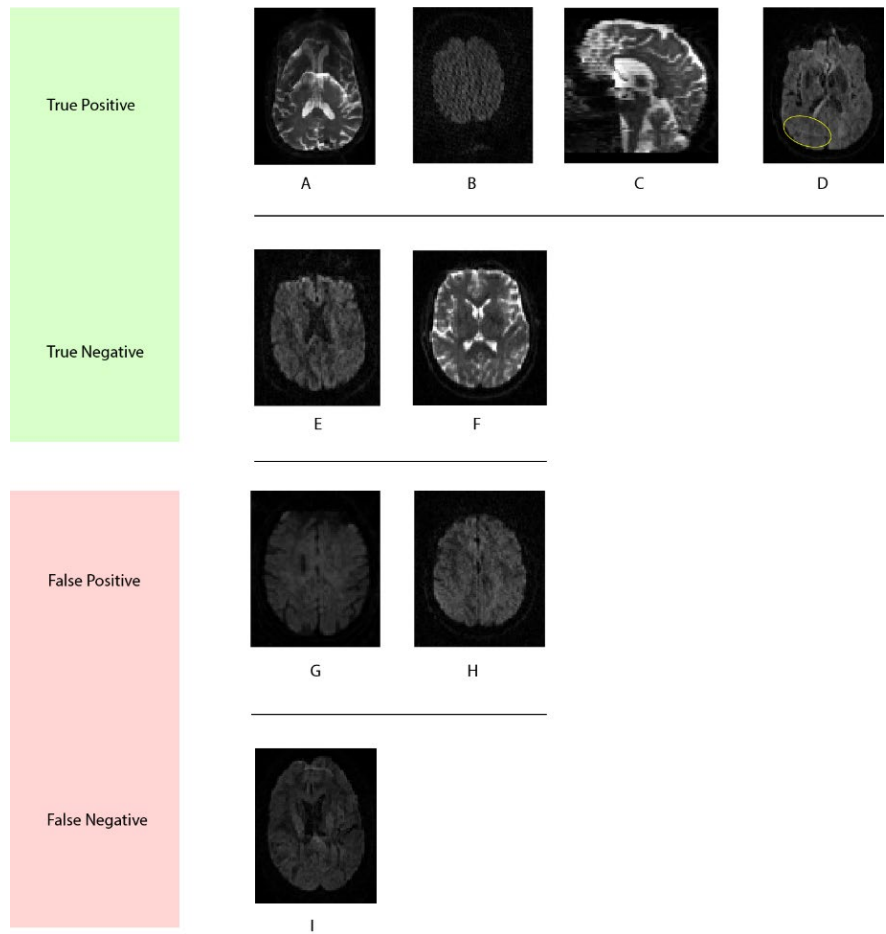

*Fig 2. Scans from the test set illustrated to demonstrate 3D-QCNet's model performance with respect to ground-truth. <u>True Positive samples</u> – **A** Ghosting artifact, **B** Herringbone artifact, **C** Motion/interslice instability artifact, **D** Faint Chemical artifact (marked in yellow). <u>True Negative samples</u> – **E** Weighted Image is noisy but is correctly marked as normal. **F** B0 image with no artifacts. <u>False Positive samples</u> – **G** Abnormal anatomy of the brain may be affecting the classifier. **H** Weighted image is noisy but there are no visible artifacts, the model may be too sensitive. <u>False Negative samples</u> – **I** Chemical shift artifact alongside instability and susceptibility.*

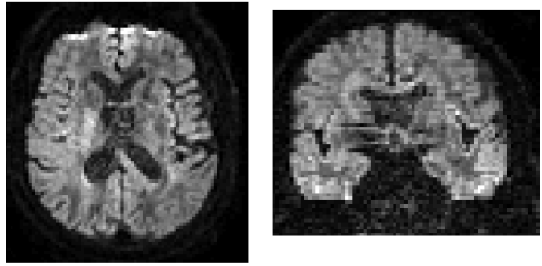

*Fig 3. These scans are from a volume that was marked as normal by our annotator but 3D-QCNet labelled it as having an artifact. Later, on closer inspection it was found to have ghosting artifacts in the ventricles and subcortical regions along with some interslice instability.*

## 4. Discussion

In this paper, we have presented a QC method to automatically detect artifacts in dMRI scans by utilizing a deep learning model trained on 3D diffusion volumes. Our overarching goal was to present users with an accurate and convenient QC toolkit that works out of the box, with minimal manual intervention needed on any practical dataset. To achieve this, it was important to demonstrate the efficacy and generalizability of our method on multiple unseen yet realistic scenarios that users may encounter in practice. Therefore, to establish the robustness of 3D-QCNet, we annotated a large amount of data - 9258 volumes from 678 subjects taken from 7 different datasets. Of these, data from 4 datasets (2347 volumes / 342 subjects), was used exclusively for testing. Significantly, the diversity, as well as the magnitude of the data we annotated, was much larger than previous automatic QC techniques. For instance, our data consisted of 4 times the number of subjects and had multiple pathologies when compared to QC-Automator which only had healthy controls. This increase in training and testing data was helped in large part by the 3D structure of our model, that only required a single label for an entire volume consisting of multiple slices, markedly reducing manual annotation time compared to 2D slice wise annotation methods. Our aim of using a vast and diverse dataset was to demonstrate the efficacy and generalizability of our method on multiple unseen yet realistic scenarios that users can expect in practice.

Having trained the deep learning model, we attempted to define the most effective way to compare results across datasets and model settings. In our discussions with human expert QC annotators, we observed that most preferred a higher artifact detection rate (higher recall) even if it meant some normal cases were marked as corrupted (lower precision). This led us to prioritize a high recall value over other evaluation metrics when comparing model results. Ideally, we would want the deep learning model to match the performance of a human rater, however, the strenuous nature of annotating a large dataset makes these ratings error prone (as seen in Fig 3). Therefore, assuming small insignificant inaccuracies inherent within the ground truth data, a recall rate close to 100 may be an unreasonable expectation. We further cannot benchmark our results to previous automatic QC studies, as in addition to not using the same test data, methods like QC-Automator provided 2D slice based results compared to our 3D volume scores. We, however, define a base recall acceptability criterion of 80% when judging the practical viability of a model. A model that captures more than 80% of the artifacts marked by the human expert is judged to be acceptable. Table 2 shows the excellent performance of 3D-QCNet. The method exceeds acceptability criterion on the validation set and its performance on the test set is even better, where it achieves a high average recall of 92 while maintaining a precision of 87. It is worth noting that these 7 datasets presented a significant challenge for validating the robustness of our method, as they contained scans from different sites, scanners, subjects and pathologies. In addition, the 4 test datasets contained a data distribution which was unseen by the model and yet the evaluation metrics on this set were found to match, and in some instances exceed the validation performance. This strongly suggests that our method can be used accurately and out of the box on most datasets regardless of their inherent characteristics. This is significant as machine learning algorithms have been found to be notoriously bad at generalizing to unseen data distributions in the field of medical imaging, and their performance can often be found lacking in practical scenarios [33].

Our method is also the first automated QC technique for dMRI, that has been tested to work on clinical datasets containing pathologies. The 3D-QCNet model generalized across a diverse set of diseases with focal abnormalities like WM hyperintensities (Dataset 4 and 5) and non-focal abnormalities (Dataset 3), or a combination thereof and spanned pathologies like autism and traumatic brain injury (Datasets 3,4 and 6). Our results demonstrate the model's ability to generalize across 4 unseen test datasets not only with pathologies, but also substantially different scanning parameters, spanning single, multishell, and state of the art multiband acquisitions as well as legacy dMRI data. This is especially demonstrated by the high recall values attained by Datasets 5 and 6 that each represented data from multiple sites and scanners and Dataset 7 which was a multishell acquisition (a scanning characteristic absent from the training data) spanning multiple b values having different levels of Signal to Noise Ratio (SNR) across them. The fact that the model was able to accurately generalize across datasets with substantial variations such as scanner and pathology differences, also suggests that our model is indeed learning to detect artifacts and not other extraneous features within the biology. Overall, 3D-QCNet's strong generalization performance allows it to be used as a convenient black-box that users can use to accurately QC any dataset, without being concerned about its intrinsic characteristics.

While in most scenarios, we expect our method to be low maintenance and work directly on datasets, as it does in Table 2, we are also mindful of the wide-ranging diversity in dMRI data that is encountered by practitioners and how it may negatively affect model performance in the future. Although we never had to use it in our experiments, as metrics easily exceeded acceptability thresholds, we provide users the ability to fine-tune on a new dataset on which the model fails to perform satisfactorily. Such a discrepancy in results might potentially be due to an unseen data distribution that the model fails to generalize upon. In such cases, selecting and manually annotating a small subset of scans from this dataset, and training the model on them enables 3D-QCNet to learn information regarding this new distribution. To simulate such a scenario, we ran the experiments in Table 3 in which we trained the model on only one dataset (Dataset 1) with a training size of 1000 volumes, and fine-tuned and tested it on 2 others (Dataset 3 and 4). The objective was to verify that performance improvement was observed on new datasets by fine tuning and to further analyze the characteristics of the ideal training subset required for it.

Results in Table 3 indicate that fine-tuning leads to a huge improvement in performance. This is encouraging, as it suggests that should the model fail to achieve acceptable results on a future dataset, annotating and training on a small subset should rectify this problem. We believe that this improvement may be because training even on a small subset of the target data allows the model to learn a normalization function. This internally transforms these new scans to match the distribution of the data the model was trained on, thereby making the deep learning model more generalizable. Ideally, we would like to avoid the scenario where our users have to resort to fine-tuning for performance improvement. Even though the data requirement for fine tuning is small, it needs a human to select, curate and annotate DTI scans for artifacts. This requires time, effort and detracts from the automated nature of our pipeline. However, in such cases, the fact that our method is based on a 3D architecture helps alleviate these problems to a certain extent. Instead of annotating 2D slices from all 3D scans (such as in 2D methods like QC Automator), users have to only assign a single label for each volume. For instance, consider the 10% fine-tuning set selected randomly from Dataset 3 for the experiment in Table 3. It had 222 volumes sourced from 102 subjects. If we were to extract 2D slices from these 222 volumes, it would result in 17,760 slices. By annotating just 222 volumes compared to 17,760 slices, we are reducing the manual overload by 98%. This significant saving of manual effort makes the prospect of fine-tuning considerably more tractable for the expert annotator.

Selecting the ideal set for fine-tuning can be a hard decision, as a tradeoff between annotation effort and model accuracy has to be accounted for. More data for training will most likely lead to a boost in metric values at the cost of more time spent labelling the data. Hence, during our experiments in Table 4, we noted how differently sourced fine-tuning sets affected model performance. We observed that in general, a set consisting of randomly chosen 200 volumes should be good enough in most cases. This set usually ranged from 10% to 30 % of our datasets. We also tried smaller datasets such as those with 100 volumes. While performance was affected, we generally found the metrics to be within the acceptability criterion. While datasets smaller than 200 volumes may produce satisfactory results occasionally, we would not recommend them. The results also suggest that model performance was agnostic to the number of subjects and the class distribution (Artifacts/Normal) within the subset of 200 volumes. This is essential since artifacts tend to be a minority class and in certain cases a randomly selected set may have very few artifactual cases. Hence a model, which does not require a curated fine-tuning set that represents both classes equitably to generate acceptable results, makes it easier for the user to select a truly random set. This is evidenced in Table 4 where Dataset 4 (which has 41% artifacts overall) produces acceptable results using a fine-tuning subset of 200 volumes containing only 6% artifacts cases.

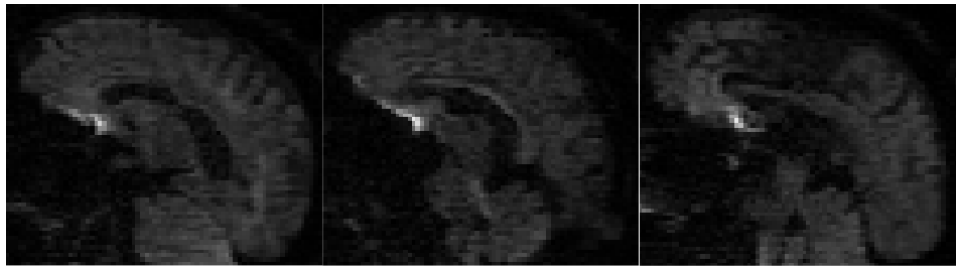

*Fig 4. This volume was marked as having an artifact by 3D-QCNet however our expert human rater (D. Parker) believes that while there is some susceptibility present, it is not enough to warrant removing the data. In such borderline cases, being able to control the model's sensitivity through probability thresholds, will allow users to finetune results based on their preferences.*

We further provide our users the ability to customize model decision by selecting flexible probability threshold values that enable the adjustment of precision-recall in a way that is more amenable to the degree of sensitivity desired by the user. Decreasing the threshold for the artifact class will lead to a more lenient criterion for assigning a volume as having an artifact. This means that more artifacts will be picked by the model, including ones that may be too faint to be deemed as such, like in Fig 4. This will lead to an overall increase in the number of ground-truth artifact cases being captured, and hence an increase in recall for the artifact class will be observed. This will often be at the cost of precision, as the lowered bar will cause many normal scans to be labelled as artifacts as well. The opposite effect, an increase in precision and decrease in recall, will be observed when the threshold is increased. Through Table 3, it can be observed how a change in threshold value can substantially alter the precision and recall metrics, in some cases making the metrics acceptable without the need for fine-tuning. An ideal value can be selected by observing the outputs on the test set using the default value of 0.5. If the precision-recall balance tilts unsatisfactorily, the user can change the threshold such that the metrics move closer to their desired values. When processing a new dataset, the user can use the default value of 0.5 (which for our 3D-QCNet seems to work well) and observe a few cases that the model labels as artifacts or normal to gauge the sensitivity of the outputs and set the threshold accordingly.

While detecting specific artifact types for each volume was beyond the scope of our problem, we manually analyzed the model predicted outputs on a significant subset of volumes to identify if there were any specific artifacts that were consistently misclassified. The various artifacts annotated in our datasets included motion-induced signal dropout, interslice instability, ghosting, chemical shift and susceptibility. 3D-QCNet was able to detect these various cases with high sensitivity and no particular type of artifact was routinely missed in testing. This is in contrast to manual QC of diffusion data, where an artifact, such as a chemical shift or a motion artifact that does not affect an entire slice, can go undetected if the rater performing QC does not diligently inspect every single slice of the data. In addition, we also observed that the model was often able to detect certain cases of artifacts that were not part of the annotated training set such as the rare Herringbone artifact (image B in fig 2), however, there sensitivity was not as high as the ones with ground truth labels.

While our method provides an accurate and convenient automated pipeline to filter artifactual DTI scans, it can be further improved upon. As 3D-QCNet only reports problematic volumes, one potential modification may involve adding the ability to identify specific slices with artifacts within the volumes. This may be useful for users who are working with smaller datasets and would prefer not to discard an entire volume if a few slices are affected. Further, since most users are only interested in weeding out the affected volumes, we do not provide fine-grained classification which identifies different artifact types. However, in rare cases where such a distinction may be desired, our method can be adapted in the future, to be trained on data with annotations for individual artifacts.

## 5. Conclusion

Quality control of dMRI data, involving the identification and removal of artifacts, is an essential step in pre-processing to ensure accurate subsequent analysis. In this paper, we proposed a deep learning based 3D classification method for artifact detection that is both accurate and convenient. It is shown to work effectively across a wide range of datasets with diverse scanner parameters and having multiple pathologies. While fine-tuning is provided as an aid to optimize it further, the generalizability of the model across 4 different test datasets suggests that it would work out of the box on future datasets. These features combined with the method's focus on minimal expert intervention, will potentially enable it to be seamlessly integrated in dMRI processing pipelines to effectively automate an essential but previously time consuming, subjective and manual process.

## Acknowledgements:

Data from various freely available, as well as clinical datasets was used for the purposes of testing the model. Only a subset of the data is free for release. The work was supported by the NIH grant - NIH R01-MH117807 (PI: Ragini Verma). We thank the following PIs and their corresponding NIH grants, to permit the use of their data for training and testing: NINDS R01 NS065980 (PI: Junghoon Kim), NINDS U01 NS086090 (PI: Ramon Diaz-Arrastia), R01-DC008871 (PI: Timothy Roberts), SPRINT (PI: Ilya Nasrallah). Data from the Philadelphia Neurodevelopmental Cohort (PNC) was also used for development and has been made freely available to the community by PIs Raquel and Ruben Gur.

# References

1. Baliyan, V., et al., *Diffusion weighted imaging: technique and applications.* World journal of radiology, 2016. **8**(9): p. 785.
2. Alexander, A.L., et al., *Diffusion tensor imaging of the brain.* Neurotherapeutics, 2007. **4**(3): p. 316-329.
3. Soares, J., et al., *A hitchhiker's guide to diffusion tensor imaging.* Frontiers in neuroscience, 2013. **7**: p. 31.
4. Le Bihan, D., et al., *Artifacts and pitfalls in diffusion MRI.* Journal of Magnetic Resonance Imaging: An Official Journal of the International Society for Magnetic Resonance in Medicine, 2006. **24**(3): p. 478-488.
5. Tournier, J.-D., S. Mori, and A. Leemans, *Diffusion tensor imaging and beyond.* Magnetic resonance in medicine, 2011. **65**(6): p. 1532.
6. Pierpaoli, C., *Artifacts in diffusion MRI.* Diffusion MRI: Theory, methods, and applications, 2010: p. 303-317.
7. Heiland, S., *From A as in Aliasing to Z as in Zipper: Artifacts in MRI.* Clinical neuroradiology, 2008. **18**(1): p. 25-36.
8. Moratal, D., et al., *k-Space tutorial: an MRI educational tool for a better understanding of k-space.* Biomedical imaging and intervention journal, 2008. **4**(1).
9. Krupa, K. and M. Bekiesińska-Figatowska, *Artifacts in magnetic resonance imaging.* Polish journal of radiology, 2015. **80**: p. 93.
10. Smith, R., R. Lange, and S. McCarthy, *Chemical shift artifact: dependence on shape and orientation of the lipid-water interface.* Radiology, 1991. **181**(1): p. 225-229.
11. Wood, M.L. and R.M. Henkelman, *MR image artifacts from periodic motion.* Medical physics, 1985. **12**(2): p. 143-151.
12. Simmons, A., et al., *Sources of intensity nonuniformity in spin echo images at 1.5 T.* Magnetic resonance in medicine, 1994. **32**(1): p. 121-128.
13. Schenck, J.F., *The role of magnetic susceptibility in magnetic resonance imaging: MRI magnetic compatibility of the first and second kinds.* Medical physics, 1996. **23**(6): p. 815-850.
14. Bammer, R., et al., *Analysis and generalized correction of the effect of spatial gradient field distortions in diffusion-weighted imaging.* Magnetic Resonance in Medicine: An Official Journal of the International Society for Magnetic Resonance in Medicine, 2003. **50**(3): p. 560-569.
15. Van Dijk, K.R., M.R. Sabuncu, and R.L. Buckner, *The influence of head motion on intrinsic functional connectivity MRI.* Neuroimage, 2012. **59**(1): p. 431-438.
16. Reuter, M., et al., *Head motion during MRI acquisition reduces gray matter volume and thickness estimates.* Neuroimage, 2015. **107**: p. 107-115.
17. Andersson, J.L., et al., *Incorporating outlier detection and replacement into a non-parametric framework for movement and distortion correction of diffusion MR images.* Neuroimage, 2016. **141**: p. 556-572.

18. Bastiani, M., et al., *Automated quality control for within and between studies diffusion MRI data using a non-parametric framework for movement and distortion correction.* NeuroImage, 2019. **184**: p. 801-812.
19. Jiang, H., et al., *DtiStudio: resource program for diffusion tensor computation and fiber bundle tracking.* Computer methods and programs in biomedicine, 2006. **81**(2): p. 106-116.
20. Oguz, I., et al., *DTIPrep: quality control of diffusion-weighted images.* Frontiers in neuroinformatics, 2014. **8**: p. 4.
21. Pierpaoli, C., et al. *TORTOISE: an integrated software package for processing of diffusion MRI data*. in *ISMRM 18th annual meeting*. 2010.
22. Liu, B., T. Zhu, and J. Zhong, *Comparison of quality control software tools for diffusion tensor imaging.* Magnetic resonance imaging, 2015. **33**(3): p. 276-285.
23. Kelly, C., et al. *Transfer learning and convolutional neural net fusion for motion artefact detection*. in *Proceedings of the Annual Meeting of the International Society for Magnetic Resonance in Medicine, Honolulu, Hawaii*. 2017.
24. Iglesias, J.E., et al. *Retrospective head motion estimation in structural brain MRI with 3D CNNs*. in *International Conference on Medical Image Computing and Computer-Assisted Intervention*. 2017. Springer.
25. Alfaro-Almagro, F., et al., *Image processing and Quality Control for the first 10,000 brain imaging datasets from UK Biobank.* Neuroimage, 2018. **166**: p. 400-424.
26. Graham, M.S., I. Drobnjak, and H. Zhang, *A supervised learning approach for diffusion MRI quality control with minimal training data.* NeuroImage, 2018. **178**: p. 668-676.
27. Samani, Z.R., et al., *QC-Automator: Deep learning-based automated quality control for diffusion mr images.* Frontiers in Neuroscience, 2019. **13**.
28. Huang, G., et al. *Densely connected convolutional networks*. in *Proceedings of the IEEE conference on computer vision and pattern recognition*. 2017.
29. Krizhevsky, A., I. Sutskever, and G.E. Hinton, *Imagenet classification with deep convolutional neural networks.* Communications of the ACM, 2017. **60**(6): p. 84-90.
30. Gu, J., et al., *Recent advances in convolutional neural networks.* Pattern Recognition, 2018. **77**: p. 354-377.
31. Ioffe, S. and C. Szegedy, *Batch normalization: Accelerating deep network training by reducing internal covariate shift.* arXiv preprint arXiv:1502.03167, 2015.
32. He, K., et al. *Deep residual learning for image recognition*. in *Proceedings of the IEEE conference on computer vision and pattern recognition*. 2016.
33. Davatzikos, C., *Machine learning in neuroimaging: Progress and challenges.* NeuroImage, 2019. **197**: p. 652.